# Quantum Circuit for Random Forest Prediction

Liliia Safina[a, *], Kamil Khadiev[a], Ilnar Zinnatullin[a], and Aliia Khadieva[a]

[a]*Kazan Federal University (KFU), Kazan, Russia*
*e-mail:[liliasafina94@gmail.com](mailto:liliasafina94@gmail.com)

**Abstract**—In this work, we present a quantum circuit for a binary classification prediction algorithm using a random forest model. The quantum prediction algorithm is presented in our previous works. We construct a circuit and implement it using qiskit tools (python module for quantum programming). One of our goals is reducing the number of basic quantum gates (elementary gates). The set of basic quantum gates which we use in this work consists of single-qubit gates and a controlled NOT gate. The number of CNOT gates in our circuit is estimated by $O(2^{n+2h+1})$, when trivial circuit decomposition techniques give $O(4^{|X|+n+h+2})$ CNOT gates, where $n$ is the number of trees in a random forest model, $h$ is a tree height and $|X|$ is the length of attributes of an input object $X$. The prediction process returns an index of the corresponding class for the input $X$.

## 1. INTRODUCTION

Nowadays, machine learning algorithms are widely used in different areas of life. Quantum machine learning algorithms are also of interest for researchers and industries. In this work, we present a quantum circuit for the binary classification prediction algorithm using random forest model [1]. Moreover, one of our goals is to estimate the number of basic quantum gates and try to reduce it. So, our prediction algorithm can be used as a part of a machine learning meta model. The model can be an ensemble model, where a random forest model is used as a basic classifier. Our goal is not only obtaining the prediction result of the random forest model, a class number, but we also want to compute (predict) the probability of $X$ belonging to the result class. To predict the probability, we use the quantum amplitude amplification and estimation algorithm [2]. The probability prediction algorithm uses the binary classification prediction algorithm and its inversion as a routine and executes them several times.

It is known that any quantum algorithm can be represented by a big unitary matrix, in other words, a multi-qubit gate. In fact, a gate-based quantum computer operates with only a predetermined set of basic gates. That means multi-qubit gates should be decomposed into circuits using elementary basic gates. Any multi-qubit gate can be decomposed by such set of basic gates. In our work, we use controlled not gate ($CNOT$) as two-qubit gate and NOT gate ($X$), rotation about y-axis gate ($Ry$), and Hadamard gate ($H$) as one-qubit gates. Also, we use $Z$-gate and its controlled version when we estimate probabilities. For the hardware implementation, the most expensive gates are two-qubit gates (in our case, $CNOT$ gates). Moreover, the computational error increases with a circuit depth, particularly, with the number of controlled gates. Therefore, one of the most important problems is to reduce their number.

Let $N$ be a number of trees in a random forest model, $O(T)$ be running time of class prediction on one tree. In the classical case, running time for the model is $O(N T)$. The quantum prediction algorithm needs $O\left(\sqrt{N}\,T\right)$ running time [1]. In this work, we construct a quantum circuit for the algorithm and estimate the number of elementary gates, actually, $CNOT$-gates.

Our quantum register looks like:
$$|anc_{mct_{rec}}\rangle|X\rangle|anc_i\rangle|i\rangle|anc_j\rangle|j\rangle|class\rangle,$$
where $|X\rangle$ is an input object, $|i\rangle$ is an index of a tree, $|j\rangle$ is an index of a node in the tree, $|class\rangle$ is a class label ($|0\rangle$ or $|1\rangle$ for the binary classification problem), $|anc_{mct_{rec}}\rangle$, $|anc_i\rangle$ and $|anc_j\rangle$ are one-qubit ancillas.

To construct the circuit, we use qiskit [3]. Qiskit is the python framework for quantum programming which allows the gate-based construction of a quantum circuit (low-level programming). In our work, for circuit implementation, we use a simulator of the real quantum device, namely, qasm_simulator of the IBMQ machine [11], and qiskit framework tools. To simplify our circuit implementation, due to limitations of computational resources of the simulator, we store $|X\rangle$ as a binary vector, where each digit (bit) is an indicator of an input object's attribute. Let $|X|$ be the length of the vector.

The quantum register has $L = 1 + |X| + n + 1 + (h - 1) + 1$ quantum bits. Note that almost all quantum algorithms can be decomposed by using $O(4^R)$ elementary gates [5], where $R$ is the number of qubits (or circuit width). In our case, $R = L$. We provide a quantum circuit and optimize it with respect to the number of CNOT gates. Our result is $O(2^{n+2h+1})\ CNOT$ gates. This result has been obtained by using different decompositions of multi-controlled not gates which are used in the tree prediction algorithm. The circuit was tested on artificially created data and was compared with results obtained by classical implementation of the algorithm.

This paper is organized as follows. In Section 2, we discuss a quantum register for the prediction problem, the quantum prediction algorithm using random forest, and a dataset for checking the correctness of the algorithm. In Section 3, the implementation of the circuit is considered. Finally, results of the optimized circuit construction are presented in Section 4.

## 2. THE PROBLEM

Note that our quantum prediction method for binary classification can be used for any ensemble model. We implement the algorithm using a random forest model on an IBMQ qasm_simulator using qiskit and test it. We also optimize the circuit by reducing the number of gates.

### *Random Forest*

Let us define our quantum prediction algorithm for binary classification problems. Let us note that the algorithm works in $O(T)$ running time. The method which we propose for getting class probabilities is based on $O\left(\sqrt{N}\right)$ times applying of the quantum amplitude amplification algorithm.

The steps of the prediction:
1. Apply the Hadamard gate to the register $|i\rangle$ to get the superposition of $2^n$ states.
2. Run the tree prediction algorithm. The gate executing this algorithm is controlled by the $|i\rangle$ register.
3. Measure the $|class\rangle$ qubit. The $|0\rangle$ result means the first class ($class_0$), the result $|1\rangle$ means the second class ($class_1$).

Due to quantum parallelism, the tree prediction works on all trees simultaneously. We construct our own data structure (a class in Python) for the random forest and call it *RandomForestForQuantumPrediction*. Attributes of the class are the next: $h_{trees}$ is the tree height (all trees have the same height and they are balanced); $trees$ are *TreeForQuantumPrediction* instances (see below); $n_{trees}$ is the number of trees in the forest.

The $RandomForestForQuantumPrediction$ has a method *predict_proba*. The method predicts probabilities of the first class and gets $X$ as an input object. The probabilities are the arithmetic mean of probabilities

$$p_{class_0} = \frac{1}{n_{trees}} \sum_{i=0}^{n_{trees}-1} p^i_{class_0},$$

where $p^i_{class_0}$ is the first-class probability on the $i$th tree.

The *TreeForQuantumPrediction* class is used to describe a tree. Its attributes are: $h$ is the tree height; $attr_{ind}$ is a list of attribute indexes of non-leaf nodes; $leaves$ is a list of angles for each leaf node. The angles are computed by the next formula:

$$leaves^i_j = arccos \sqrt{p_{j,class_0}{}^i},$$

where $i$ is the number of a tree and $j$ is the number of a leaf node in the $i$th tree.

We construct a sample of trees. Due to limitations of quantum computing resources of the simulator, the sample is simple enough, so that we can easily check the result. Let us present our random forest in Figure 1. The numbers in nodes of the trees are attribute indexes, the numbers in brackets near the nodes are the node indexes. The numbers below the leaves are the $class_0$ probabilities.

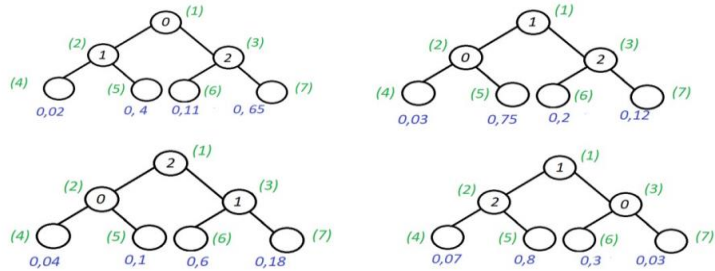

*Figure 1. The Random Forest*

*Test Dataset*

Note that the $|X\rangle$ register is a binary vector. Simulation process on qiskit takes a long time. So, we test the quantum circuit on artificially created data with short vector $X$. We set $|X| = 3$. Prediction results obtained by classical implementation which uses the random forest given in Figure 1 are shown in Table 1.

*Table 1. Prediction Results*

| $X$ | $p_{class_1}$ |
|---|---|
| 000 | 0.04 |
| 001 | 0.3625 |
| 010 | 0.235 |
| 011 | 0.25 |
| 100 | 0.2575 |
| 101 | 0.7 |
| 110 | 0.11 |
| 111 | 0.245 |

## 3. IMPLEMENTATION

Qiskit is a framework for quantum programming on IBM quantum computers based on Python language. It allows the execution of quantum algorithms by construction of quantum circuits. Today, quantum computers are limited by memory resources and not resistant to noise. So, one of the significant goals of quantum computing is minimization of depth of the circuits implementing the algorithms, which means that we should use as few quantum gates as possible.

Any quantum computer operates with a predetermined set of basic gates. The bigger constructions are transpiled into the circuit of these gates. In work [4], authors prove universality of a set of gates that consists of all one-bit quantum gates and the two-bit exclusive-or gate ($CNOT$ gate). There was also shown that $O(4^c)$ $CNOT$-gates are enough for decomposition of any quantum circuit in common case [4], where $c$ is the number of qubits.

We construct circuits for the random forest prediction process and count elementary gates. We use a qasm simulator of a real machine to run the circuits. Qiskit provides its own realization of multi-qubit gates. We can obtain the decomposition results using the *decompose* method and count elementary gates by the *count_ops* method. Some methods in the prediction process are controlled by one or more qubits. For these cases, we use the *control* method provided by qiskit. Also, there is a multi-controlled not gate ($mct$) in qiskit. First, we use them to construct our circuit. All circuits presented in this work use the trees presented in Figure 1 for $X$ of size 3.

Let us present our prediction circuit given in Figure 2.

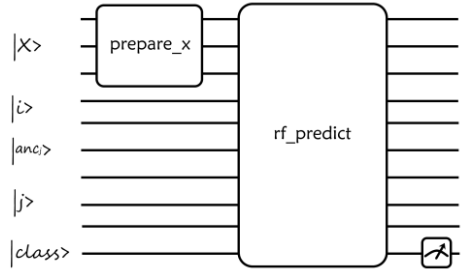

*Figure 2. RF prediction*

The *prepare_x* is a method that obtains the state $|X\rangle$ corresponding to the input vector $X$ by using *NOT*-gates (*X*-gates). Let us present the *rf_predict* method.

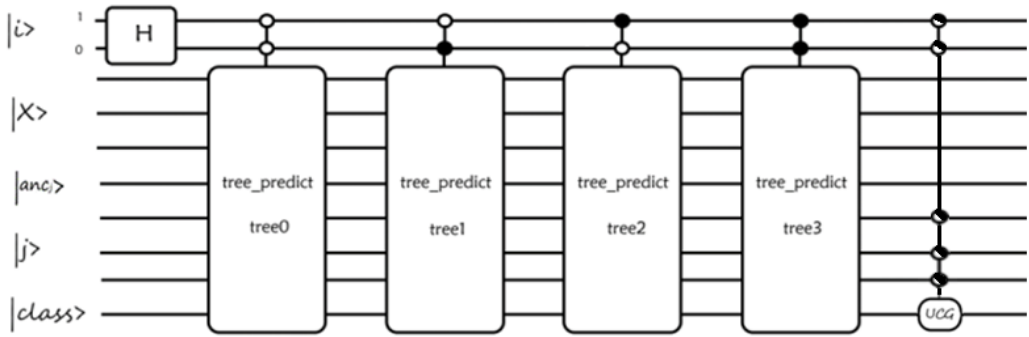

*Figure 3. The method rf_predict*

In Figure 3, we show the circuit of the prediction process using the model presented in Figure 1. The first step is to get the equally distributed superposition of all trees. Then, we run the prediction process on the trees. At this step, the *tree_predict* operator is controlled by the $|i\rangle$ register, where the *tree_predict* operator changes the $|j\rangle$ register according to the value of $X_{ind_j}$. It is an attribute (one bit) of $X$ in a place $ind_j$, and it depends on itself. Obviously, the register cannot be target and controller at the same time, so we use $anc_j$ qubit. The *tree_predict* method checks all the node indexes. The index $ind_j$ is a given parameter of the $j$th tree. If $X_{ind_j} = 0$, then the register $|j\rangle$ becomes $|2j\rangle$, and it becomes $|2j + 1\rangle$, otherwise. To get $|2j\rangle$, we use a few $SWAP$s. The number of $SWAP$s is $O(h)$. Qiskit has a few methods to multiply 2 numbers, but the methods need extra qubits and they are expensive to decompose them. For our problem $SWAPS$ are enough to multiply the number $j$ by 2. To decompose a $SWAP$-gate, we use 3 $CNOT$-gates. To get $|2j + 1\rangle$, we use also $SWAP$-gates and one $NOT$-gate for the last qubit. The *tree_predict* method stops when the $|j\rangle$ becomes a leaf. When all trees finish the prediction process, we use uniformly controlled gate (UCG) [5] to rotate the $|class\rangle$ qubit. The rotation is controlled by $|i\rangle$ and $|j\rangle$. Each tree and all leaves have their own rotation angles, they are computed using the first-class probabilities (given parameters).

Then, we re-enumerate the leaves of the trees. That allows us to reduce the number of qubits and gates. The new indexes are presented in Figure 4. The leaves are again enumerated starting from 0. When the register $|j\rangle$ becomes a leaf, its index of the leaf in the tree is $2j + X_{ind_j}$ if we don't take into account the highest bit in the binary representation of the leaves indexes. Now, to store the $|j\rangle$ register, we use only $h - 1$ qubits. Due to the actual highest bit of the binary representations of leaves indexes is equal to 1. So, we apply a few $SWAP$s and one subtraction operation "$-1$" (if $X_{ind_j} = 0$), because the actual highest bit after the $SWAP$s becomes the lowest bit. The operation $-1$ is also implemented by using UCG. We call the method $dec\_ucg$.

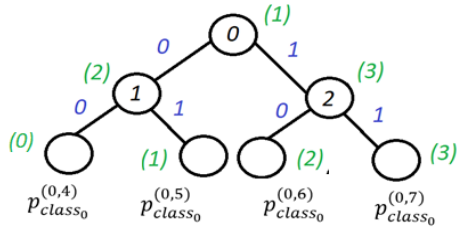

Figure 4. The new tree data structure

At the next step, we add one more ancilla for $|i\rangle$. Now, the tree prediction process is controlled by one ancilla. Controlled functions are decomposed by using a sequence of $mct$ gates. Note that $mct$ is not an elementary gate and is decomposed by using a big amount of basic gates [10]. The new prediction process is presented in Figure 5.

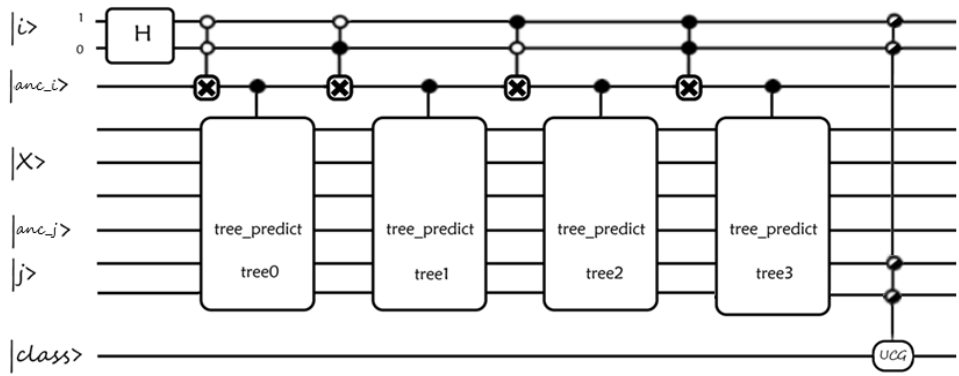

Figure 5. The rf_predict circuit

After that, we rewrite the *tree_predict* method, and augment it by a control qubit. Now, it is called the *tree_predict_controlled* method. The approach allows us to reduce the number of elementary gates, because our $mct$-implementations are cheaper than original versions. Let us demonstrate numbers of elementary gates in decompositions of $mct$ in Table 2.

In Table 2 the label $U$ means a single-qubit gate. The $mct\_ucg$ is our implementation of $mct$. It is based on UCG idea. The qiskit gate *mct (mode='recursion')* requires one more ancilla. The table demonstrates that $mct\_ucg$-gate is the cheapest one if the number of controlled qubits is less than 8. In the general case, a random forest model uses trees with heights in a range 5-10. So, for checking the node indexes, we use $mct\_ucg$. For checking the tree indexes, we add condition: if the number of trees is more than $2^7$, then we use *mct (mode='recursion')* and $|anc_{mct_{rec}}\rangle$ as an ancilla, and $mct\_ucg$ without ancilla, otherwise.

*Table 2. MCT-decompositions*

| The controlled qubits number | mct | | mct (mode='recursion') | | mct_ucg | |
|---|---|---|---|---|---|---|
| | U | CX | U | CX | U | CX |
| 2 | 9 | 6 | 9 | 6 | 4 | 3 |
| 3 | 17 | 14 | 17 | 14 | 8 | 7 |
| 4 | 69 | 36 | 69 | 36 | 16 | 15 |
| 5 | 95 | 92 | 68 | 56 | 32 | 31 |
| 6 | 191 | 188 | 172 | 100 | 64 | 63 |
| 7 | 383 | 380 | 276 | 144 | 128 | 127 |
| 8 | 767 | 764 | 274 | 184 | 256 | 255 |
| 9 | 1535 | 1532 | 272 | 224 | 512 | 511 |

Actually, a lot of operations after using controlled qubits in the *tree_predict_controlled* method need only 2-3 controlled qubits. Then, we use one more multi-controlled $NOT$-gate which is $rcccx$ [6], 3-controlled $NOT$-gate. Its decomposition is presented in a work [7]. The control qubits are $anc_i$, $anc_j$ and one more for $CNOT$ or controlled by $X_{ind_j}$. The gate is used in $SWAPS$ and in the +1 operation ($inc$). The $rcccx$-gates need 6 $CNOT$-gates and 12 $U$-gates (vs. 7 $CNOT$s and 8 $U$-gates in the cheapest $mct\_ucg$). We remind that $CNOT$-gate is more expensive than a single-qubit gate for physical implementation.

*Gate Counting*

Let us remind, that $h$ is the tree height, n is the number of qubits to store the tree indexes.

For each tree we iteratively set $anc_i = 1$, and then, we run the *tree_predict_controlled* method. The steps take $O(2^n)$ iterations. The *tree_predict_controlled* makes $O(2^h)$ iterations. Each iteration is a few $mct\_ucg$ and $rcccx$. The most expensive operations of them are $mct\_ucg$ gates. This gate requires $O(2^h)$ $CNOT$s [5]. We need to set all ancillae to be equal to 0 again after each iteration. So, we have $O(2^n \times 2^h \times 2^h \times 2) = O(2^{n+2h+1})$. Remind that the number of elementary gates to decompose almost all operators on R qubits is $O(4^R)$, in our case $R = |X| + n + h + 2$.

*Amplitude Amplification and Estimation*

After the circuit optimization, we tested the quantum random forest prediction using amplitude estimation [8]. We use the simplest estimation algorithm from [2] which is an algorithm $QSearch$. The idea of it is close to Grover's search algorithm [9]. The algorithm iteratively repeats $Q = AS_0A^{-1}S_\chi$, where $A$ is a quantum algorithm, $S_0$ changes the sign of a zero state, $A^{-1}$ is inversion of $A$, and $S_\chi$ is an oracle. The expected numbers of $Q$ iteration to find a $good$ element is equal to $\frac{1}{\sqrt{p}}$, where $p$ is the probability of the $good$ element.

In our case, $A$ is the random forest prediction algorithm, $p = p_{class_0}$, a $good$ element is the $class_0$ state. We count the number of $Q$ repeatings while a measured class is not $class_0$. If we know the number of iterations $k$, we can compute an approximate value $p_{class_0} = \left(\sin\frac{\pi}{4k}\right)^2$.

To implement $S_0$ and $S_\chi$, we use $NOT$s, $Z$-gates and controlled $Z$-gates. To inverse the $rf\_predict$ function, we use the $inverse$ method of qiskit.

## 4. RESULTS

We run the final circuit 100 times on a qasm simulator and measure the $|class\rangle$ register. The values $X \in \{000, 011, 101, 110\}$ are the most informative. Other predictions also work correctly. We predict result classes using the classical $predict\_proba$ method and our quantum prediction algorithm. The results are in Tables 3-4.

*Table 1. Predict_proba results*

| $X$ | $p_{class_0}$ |
|---|---|
| 000 | 4% |
| 011 | 25% |
| 101 | 70% |
| 110 | 11% |

*Table 2. Quantum prediction results*

| $X$ | $class_0$ ($|0\rangle$) | $class_1$($|1\rangle$) |
|---|---|---|
| 000 | 6 | 94 |
| 011 | 31 | 69 |
| 101 | 68 | 32 |
| 110 | 17 | 83 |

The tables show that the quantum prediction works correctly.
The amplitude estimation results are presented in Table 5.

*Table 5. Estimation results*

| $X$ | The number of launches of $Q$ | Estimation results |
|---|---|---|
| 000 | 3 | 0.07 |
| 011 | 1 | 0.5 |
| 101 | 0 | 1 |
| 110 | 1 | 0.5 |

The number of launches cannot be non-integer, so, we cannot get more accurate results for these probabilities. If actual probabilities are small enough, then the number of launches increases, and so, we can obtain more accurate results.

## 5. CONCLUSIONS

We constructed quantum circuits for the binary classification problem using a random forest model. Moreover, we estimated the number of elementary gates in the circuit and optimized it by reducing the number of gates. To reduce the circuit depth, we transformed our methods which lie in a basis of the prediction process, and we also implemented multi-controlled $NOT$-gates using our optimization technique. This implementation is based on a uniformly controlled gate. We also add a control qubit inside our prediction functions.

These steps allow us to get the number of $CNOT$s of an order $O(2^{n+2h+1})$. For the case of using of the general decomposition, this number would be $O(4^{|X|+n+h+2})$.

The implementation is in [12].

## FUNDING

The study was funded by the subsidy allocated to Kazan Federal University for the state assignment in the sphere of scientific activities, project No. 0671-2020-0065.

## CONFLICT OF INTEREST

The authors declare that they have no conflicts of interest.

# REFERENCES

1. K.R. Khadiev and L.I. Safina, "The quantum version of prediction for binary classification problem by ensemble methods" International Conference on Micro-and Nano-Electronics 2021. – SPIE, pp. 595-603 (2022).

2. G. Brassard, P. Høyer, M. Mosca, and A. Tapp, "Quantum amplitude amplification and estimation," Contemporary Mathematics 305, 53–74 (2002).

3. "Qiskit 0.44 documentation", url=https://qiskit.org/documentation/.

4. A. Barenco, C. H. Bennett, R. Cleve, D. P. DiVincenzo, N. Margolus, P. Shor, T. Sleator, J. A. Smolin, and H. Weinfurter, "Elementary gates for quantum computation," Physical Review A 52, 3457–3467 (nov 1995)

5. D. M. Möttönen and J.J. Vartiainen," Decompositions of general quantum gates", Trends in Quantum Computing Research, pp. 149 (2006).

6. "CLASSRC3XGate",

url= https://qiskit.org/documentation/stable/0.26/stubs/qiskit.circuit.library.RC3XGate.html

7. D. Maslov, "On the advantages of using relative phase Toffolis with an application tomultiple control Toffoli optimization", arXiv preprint arXiv:1508.03273.

8. K. Khadiev, L. Safina, "The quantum version of random forest model for binary classification problem", CEUR Workshop Proc., pp. 30-35 (2021).

9. L. K. Grover, "A fast quantum mechanical algorithm for database search ", Proceedings of the twenty-eighth annual ACM symposium on Theory of computing, pp. 212-219 (1996).

10. Miller D. M., Wille R., Sasanian Z. Elementary quantum gate realizations for multiple-control Toffoli gates //2011 41st IEEE international symposium on multiple-valued logic. – IEEE, 2011. – C. 288-293.

11. IBMQ, url=" https://quantum-computing.ibm.com"

12. https://github.com/LiliiaSafina/q_rf_predict.git